# Comments on Failures of van der Waals' Equation at the Gas–Liquid Critical Point, L. V. Woodcock, International Journal of Thermophysics (2018) 39:120

I.H. Umirzakov

Institute of Thermophysics, Novosibirsk, Russia
cluster125@gmail.com

**Abstract**
   These comments are a response to the discussion presented in the above paper concerning the "New comment on Gibbs Density Surface of Fluid Argon: Revised Critical Parameters" by Umirzakov. Here we show that: Woodcock's results obtained for the dependencies for the isochoric heat capacity, excess Gibbs energy and coexisting difference functional of argon, and coexisting densities of liquid and vapor of the van der Waals fluid and presented in all Figures are incorrect; his Table includes incorrect values of coexisting difference functional; his paper includes many incorrect equations, mathematical and logical errors and physically incorrect assertions concerning the temperature dependences of the isochoric heat capacity and entropy of real fluids; most of the his conclusions are based on the above errors, incorrect data, incorrect comparisons and incorrect dependencies; and most of his conclusions are invalid. We also show that the van der Waals equation of state quantitatively describes the dependencies of saturation pressure on vapor density and temperature near critical point, and the equation of state can describe qualitatively the reduced excess Gibbs energy, rigidity and densities of coexisting liquid and vapor of argon, including the region near critical point.

## 1. Introduction

   Our comments are a response to a discussion of the article "New comment on Gibbs Density Surface of Fluid Argon: Revised Critical Parameters" by Umirzakov [1] held in the paper [2]. Here we show that 1) the dependencies presented in all Figures in [2] for the isochoric heat capacity, excess Gibbs energy and coexisting difference functional of argon, and coexisting densities of liquid and vapor of the van der Waals fluid are incorrect; 2) Table 1 [2] includes incorrect values of coexisting difference functional; 3) the paper [2] includes many incorrect equations, mathematical and logical errors, incorrect comparisons and physically incorrect assertions concerning the temperature dependences of the isochoric heat capacity and entropy of the real fluids; 4) most of the conclusions in [2] are based on the above errors, incorrect data, incorrect comparisons and incorrect dependencies; and most of the conclusions in [2] are invalid; 5) the van der Waals equation of state quantitatively describes the dependencies of saturation pressure on saturated vapor density and temperature near critical point; and 6) the equation of state can describe qualitatively the excess Gibbs energy, rigidity and densities of coexisting liquid and vapor of argon (real fluid), including the region near critical point.

## 2. Comments

1. Let us consider the first assertion discussed in [2]: "In contrast to the conjecture [1] there is no reliable experimental evidence to doubt the existence of a single critical point," citing the Sengers and Anisimov comment [2] based upon historic evidence from divergent isochoric heat capacity $C_V$ measurements at the critical temperature ($T_c$)".

   The first part of the assertion ("In contrast to the conjecture [1] there is no reliable experimental evidence to doubt the existence of a single critical point") was quoted from [3] in [1], but there was not the rest part of the assertion ("citing the Sengers and Anisimov comment [2] based upon historic evidence from divergent isochoric heat capacity $C_V$ measurements at the critical temperature ($T_c$)") in [1]. The first part of the assertion means that there is no reliable experimental evidence to doubt the existence of a single critical point and this is in contrast to the conjecture of [2] and nothing more. So, the first assertion discussed in [2] is an incorrect assertion from [1], while a correct assertion from [1] is:

   Assertion 1. "In contrast to the conjecture [1] there is no reliable experimental evidence to doubt the existence of a single critical point".

   From logical point of view, it is clear that an experimental proof of the existence of two or more critical points or the existence of a critical line will be the proof of the incorrectness of the Assertion 1. However, such experimental proof was not presented in [2]. Moreover, one can see from [2] that there are no other proofs in [2] for the Assertion 1 to be incorrect.

   It is evident that the Assertion 1 does not mean that Anisimov and Sengers divergent $C_V$ at $T_c$ is wrong. Therefore, the conclusions of [2] that "if Umirzakov's first assertion were to be right, Anisimov and Sengers divergent $C_V$ at $T_c$ would have to be wrong. In fact, neither of the assertions will withstand scientific scrutiny" have no sense.

2. The second assertion discussed in [2] is "(quote) "… to prove that the existence of a single critical point of a fluid described by van der Waals equation of state (VDW-EOS) is not a hypothesis and is a consequence of the thermodynamic conditions of liquid–vapor phase equilibrium."

   One can see from [1] that the quote in the second assertion is incorrect and a correct assertion from [1] is:

   Assertion 2. "We prove that the existence of a single critical point of a fluid described by van der Waals equation of state (VDW-EOS) is not a hypothesis and is a consequence of the thermodynamic conditions of liquid–vapor phase equilibrium".

   It is easy to see reading [2] that there is no proof in [2] that the existence of a single critical point of the fluid described by the van der Waals equation of state is hypothetical and the existence of a single critical point of the van der Waals fluid is not a consequence of the thermodynamic conditions of liquid–vapor phase equilibrium. So, there is no proof in [2] that the Assertion 2 is incorrect.

3. One can see that the van der Waals equation of state [4] alone was considered in [1] and all conclusions of [1] concern the van der Waals fluid. There is no statement or assumption in [1] that the van der Waals equation of state describes quantitatively the thermodynamic properties of the real fluids. It is evident that the statements of [1] that "there is no reliable experimental evidence to doubt the existence of a single critical point" and "the existence of a single critical point of a fluid described by the van der Waals equation of state (VDW-EOS) is not a hypothesis and is a consequence of the thermodynamic conditions of liquid–vapor phase equilibrium" do not mean that the van der Waals equation of state describes quantitatively the thermodynamic properties of the real fluids (for example, of argon). It is also evident that the proof that the van der Waals equation of state cannot describe quantitatively the thermodynamic properties of the real fluids does not mean that the above statements of [1] are incorrect. There is no proof in [2] that the above and other statements of [1] are incorrect.

One can see from the above comments that there is the lack of logic in the reasoning of Woodcock in [2].

4. According to [2], "it was incorrectly asserted that van der Waals equation "proves" the existence of a scaling singularity with a divergent isochoric heat capacity ($C_V$)" in [1]. One can easily see from [1] that there is no assertion that van der Waals equation proves the existence of a scaling singularity with a divergent isochoric heat capacity in [1].

5. The Van der Waals' equation of state [4]

$$p(T,V) = RT/(V-b) - a/V^2, \qquad (1)$$

where $p$ is pressure, $T$ is temperature, $V$ is molar volume, $R$ is the molar gas constant, $a$ and $b$ are positive constants, predicts that $C_V$ is equal to that of the ideal gas $C_{V,ig}$ [5-7,9-11]. $C_{V,ig}$ is a function of temperature alone; it may be independent of temperature; $C_{V,ig}$ is independent of density (volume); $C_{V,ig}$ of the atomic fluids differs from that of molecular fluids; $C_{V,ig}$ of molecular fluids depends, particularly, on the spatial structure and masses of the atoms consisting the molecule as well as interactions between the atoms; and $C_{V,ig}$ of various molecular fluids can differ from each other [6]. So, the statements in [2] that: a) "Van der Waals' equation … erroneously predicts, for instance, that $C_V$ is a constant for all fluid states"; b) "van der Waals equation predicts the same heat capacity ($3R/2$) for all thermodynamic states of all fluids"; and c) "Equation 1 … predicts that all fluids have a constant $C_V$, i.e. equal to that of the ideal gas ($3R/2$)" are incorrect.

6. The entropy (per molecule) of the ideal gas $S_{ig}$ consisting of molecules depends on the temperature. For example, entropy of the molecule consisting of two different atoms with masses $m_1$ and $m_2$ which is approximately equal to [6]

$$S_{ig} = \left(\frac{\partial}{\partial T}\left[kT\ln\left(ev\left(\frac{MkT}{2\pi\hbar^2}\right)^{3/2}\sum_{\alpha=0}^{\infty}\exp\left(-\frac{\hbar\Omega}{kT}(\alpha+1/2)\right)\cdot\sum_{\beta=0}^{\infty}(2\beta+1)\exp\left(\frac{-\hbar^2\beta(\beta+1)}{2IkT}\right)\right)\right]\right)_v,$$

where $M = m_1 + m_2$ is the mass of the molecule, $k$ is the Boltzmann's constant, $\hbar$ is the Planck's constant, $v$ is the volume per molecule, $I$ is the moment of the inertia of the molecule, $\Omega$ is the frequency of linear (harmonic) oscillations of the molecule, depends on temperature.

Therefore, the statement in [2] that "Entropy of the ideal gas is independent of temperature at constant volume" is incorrect.

7. The entropy of the ideal gas $S_{ig}$ depends on temperature in general case. So, it is clear that the equality $\Delta S_{ig} = Q_{rev}/T$, were $\Delta S_{ig}$ is the change of entropy of the ideal gas, may be valid if heat $Q_{rev}$ is added reversibly to a real fluid at constant volume $V$. Therefore, the assertion in [2] that "we know that since entropy is a state function, and by definition, $\Delta S = Q_{rev}/T$ (where $Q_{rev}$ is reversible heat added), $S^*$ must increase to some extent with $T$ if heat is added reversibly to a real fluid at constant $V$" could be incorrect.

8. It is easy to establish from Eqs. 1-6 [2] that Eqs. 2-6 [2] for excess state functions relative to an ideal gas ($V \to \infty$) are incorrect and they must be replaced by

$$A^* = -\int_{\infty}^{V} P^* dV = -RT \ln[(V-b)/V] - a/V,$$

$$S^* = \int_{\infty}^{V} (\partial P^*/\partial T)_V dV = R \ln[(V-b)/V],$$

$$U^* = A^* + TS^* = -a/V,$$

$$H^* = U^* + P^*V = RTb/(V-b) - 2a/V,$$

$$G^* = H^* - TS^* = RT[b/(V-b) - \ln(1-b/V)] - 2a/V,$$

where $P^*(T,V) \equiv P(T,V) - RT/V = RTb/V(V-b) - a/V^2$.

9. It is easy to establish from Eq. 8 [2] and $\rho = 1/V$ that Eq. 9 [2] is incorrect and it must be replaced by

$$"H"(y \text{ in ref. } 4) = b^4/a \cdot (\partial p/\partial V)_T = 2\omega b^4/aV^3 + b^4(\partial \omega/\partial \rho)_T/aV^4.$$

10. The critical temperature $T_c$ must have a positive value [4-11]. So, the statement in [2] that "The density difference, $\Delta F(y) = F_{liq}(y) - F_{gas}(y)$ (see figure 1 of Ref. [4]) does not go to zero $T_c$ in the case of a real fluid" has no sense.

11. According to Fig. 2 [1], the first and second partial derivatives of pressure with respect to volume at constant temperature go to zero in the limit $y \to 0+$, which means that $T_c$ is reached from the side of low temperatures (see Eq. 6 and Fig. 1 [1]). According to [2], the rigidity $\omega$ and its density derivatives go to zero for real gas and liquid states at $T_c$, $p_c$, if $T_c$ is reached from the side of high temperatures. Therefore, the statements in [2] that "Figure 2 in Ref. [4], showing that these two derivatives go to zero when $y=1$, does not prove anything because ω and its density derivatives all go to zero for real gas and liquid states at $T_c$, $p_c$. This is illustrated in Fig. 5 for the behavior of the rigidity of argon along the critical isotherm, compared to the prediction of van der Waals equation" are incorrect.

12. As one can see from Figs. 1 and 2 [1], $\Delta F = F_L - F_V$, $H_L$, $H_V$, $G_L$ and $G_V$ vanish at $y = 0$, therefore, the statements in [2] that "the fact that "$\Delta F$", "$H$" and "$G$" go to zero at $y=1$ for both coexisting gas and liquid in figures 1 and 2 of Ref. [4], respectively, does not prove anything about criticality of real fluids" have no sense.

13. It was shown earlier in [12] that the Van der Waals equation of state near critical point can be presented in an asymptotic form of the equation of state of scaling theory. So, the assertion in [2] that "Van der Waals equation, however, is inconsistent with the universal scaling singularity concept" is incorrect.

14. We proved in [1] that van der Waals fluid has only one critical point. Therefore, the statement in [2] that "Ref. [1] proves nothing more than van der Waals' equation has a singularity with two vanishing derivatives" is incorrect if the singularity does not mean that there is only one critical point.

15. The ability of the van der Waals equation of state to describe the thermodynamic properties of real fluid (e.g. argon) was not considered in [1]. The fact that the van der Waals equation of state cannot describe quantitatively the thermodynamic properties of the real fluids, including argon, was earlier established by many authors [5-12]. So, the statement in [2] that "state functions of van der Waal's equation fail to describe the thermodynamic properties of low-temperature gases, liquids and gas-liquid coexistence" is not a new insight into the science or physics.

16. Many conclusions in [2] are based on the fact that the van der Waals equation of state cannot describe quantitatively the thermodynamic properties of the real fluids. This fact does not prove the statements in [2] such as "The conclusion that there is no "critical point" singularity on Gibbs density surface remains scientifically sound", "the conclusion in Ref. [1], i.e., that there is no critical point singularity with scaling properties on Gibbs density surface still holds true", and "Van der Waals hypothetical singular critical point is based upon a common misconception that van der Waals equation represents physical reality of fluids".

17. According to [2], "Explicitly built into the equation is an incorrect a priori assumption of continuity of liquid and gaseous states." One can see from the detailed consideration of [2] that there is no proof of the incorrectness of a priori assumption of continuity of liquid and gaseous states in [2].

18. It is easy to see that the van der Waals equation of state defines an exact position of the critical point 1) on the (temperature, pressure)-thermodynamic plane if the coefficients of the equation are defined via ($T_c, p_c$) using the relations $a = 27R^2T_c^2/64p_c$ and $b = RT_c/8p_c$; 2) on the (density, pressure)-plane if the coefficients are defined via ($V_c, p_c$) using the relations $a = 3p_cV_c^2$ and $b = V_c/3$; and 3) on the (density, temperature)- plane if the coefficients are defined via ($V_c, T_c$) using the relations $a = 9RT_cV_c/8$ and $b = V_c/3$. So, the statement in [2] that "the liquid–gas critical point is not a property that the van der Waals equation can make any statements about" is incorrect.

19. According to [13], which was discussed in [1,3], pressure $p(T,\rho)$ of "meso-phase" is a linear function of density:

$$p(T,\rho) = p_0(T) + \omega(T)\rho, \qquad (2)$$

where $\omega(T)$ is rigidity and $\rho_B(T) \leq \rho \leq \rho_A(T)$. The thermodynamic relation $(\partial C_V/\partial v)_T = T(\partial^2 p/\partial T^2)_V$ [7], where $v = m/\rho$ and $m$ is the molar mass, gives

$$C_V(T,\rho) = Tm \cdot d^2p_0/dT^2 \cdot [1/\rho - 1/\rho_B(T)] - Tm \cdot d^2\omega/dT^2 \cdot \ln[\rho/\rho_B(T)] + C_V(T,\rho_B(T)). \qquad (3)$$

We conclude from Eq. 3 that $C_V$ does not depend on density, if $d^2 p_0 / dT^2 = 0$ and $d^2 \omega / dT^2 = 0$, and $C_V$ is not a linear function of density, if $d^2 p_0 / dT^2 \neq 0$ or (and) $d^2 \omega / dT^2 \neq 0$. As one can see from Fig. 1 a) in [2], the critical isotherm of the isochoric heat capacity $C_V$ in "mesophase" decreases linearly with increasing density. Hence, the isotherm of the isochoric heat capacity presented in Fig. 1 a) in [2] is incorrect.

20. The density dependence of the isochoric heat capacity of argon calculated using the fundamental equation of state (EOS) [14], which is used in the NIST database [15], is presented in Fig. 1. As on can see from Fig. 1, there is no density interval where the isochoric heat capacity decreases linearly on density. This is another proof of the incorrectness of the dependence presented in Fig. 1a) [2].

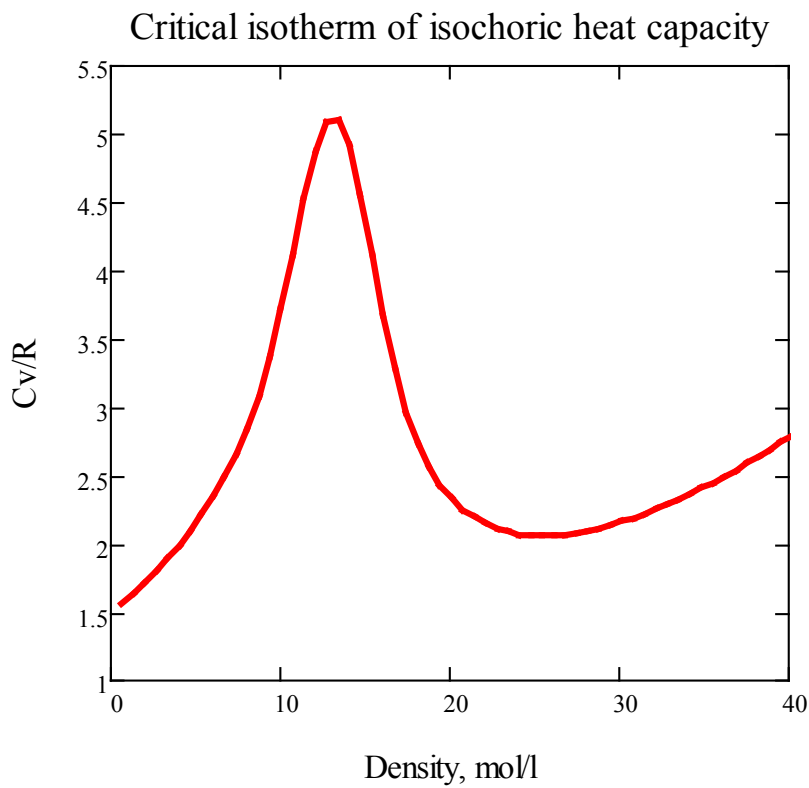

Fig. 1. The critical isotherm of the isochoric heat capacity of argon obtained using EOS [14]

21. It was shown earlier in [16] that: (1) the expressions for the isochoric and isobaric ($C_P$) heat capacities of liquid and gas, coexisting in phase equilibrium, the heat capacities at saturation of liquid and gas ($C_\sigma$) and the heat capacity ($C_\lambda$) used in Woodcock's article [17] are incorrect; (2) the conclusions of the article based on the comparison of the incorrect $C_V$, $C_P$, $C_\sigma$ and $C_\lambda$ with experimental data are also incorrect; (3) the lever rule used in [17] cannot be used to define $C_V$ and $C_P$ in the two-phase coexistence region; (4) a correct expression for the isochoric heat capacity describes the experimental data well; (5) there is no misinterpretation of near-critical gas–liquid heat capacity measurements in the two-phase coexistence region; (6) there are no proofs in the article that: (a) the divergence of $C_V$ is apparent; (b) it has not been established experimentally that the thermodynamic properties of fluids satisfy scaling laws with universal critical exponents asymptotically close to a single critical point of the vapor–liquid phase transition; and (c) there is no singular critical point on Gibbs density surface. Many mathematical and logical errors were also found in [17]. The continuous isochore of $C_V$ for $Ar$ was obtained in [17] by using the incorrect dependence of $C_V$ on temperature and density. The comparison of Fig. 1b [2] with Fig. 1b [17] shows that the isochores of argon in them are same.

The dependence of the isochoric heat capacity $C_V$ of argon along an isochore in the middle of the critical divide (density $13.3\, mol \cdot l^{-1}$) presented in Fig. 1 b) [2] which has no discontinuity is incorrect because, according to experiments, $C_V$ along an isochore must have a discontinuity when the isochore of $C_V$ passes through the coexistence line [3, 16, 18, 19].

22. The reduced excess Gibbs energy of argon relative to $G_c^* = G^*\big|_{T_c, P_c}$ for the critical isotherm obtained using EOS [14] is presented in Fig. 2 (red squares). One can see that Fig. 2b [2] is incorrect. The comparison of Fig. 2 with Fig. 2 a) [2] shows that the van der Waals equation qualitatively describes the excess Gibbs energy of argon in the critical region. Therefore, the comparison of the dependencies presented in Figs. 2 a) and b) [2] is incorrect, and the statements of [2] that "Gibbs energy of argon, taken from the NIST thermophysical property tables [7], by comparison shows that the van der Waals equation completely misses the essential behavior, especially in the vicinity of the critical point", "the absurd minimum $G^*$ at $\rho \sim 20\, mol \cdot l^{-1}$ and subsequent increase for the hypothetical van der Waals liquid are consequences of $V < b$ in Eq. 1 at this density", and "it is evident from Fig. 2a, b that the van der Waals equation fails to describe even qualitatively the thermodynamic properties of gas–liquid coexistence in the critical region" are incorrect. Fig. 2 shows that the minimum of $G^*$ at $\rho \sim 20\, mol \cdot l^{-1}$ for the van der Waals fluid is not absurd and the van der Waals equation can describe quantitatively the excess Gibbs energy of argon in the critical region.

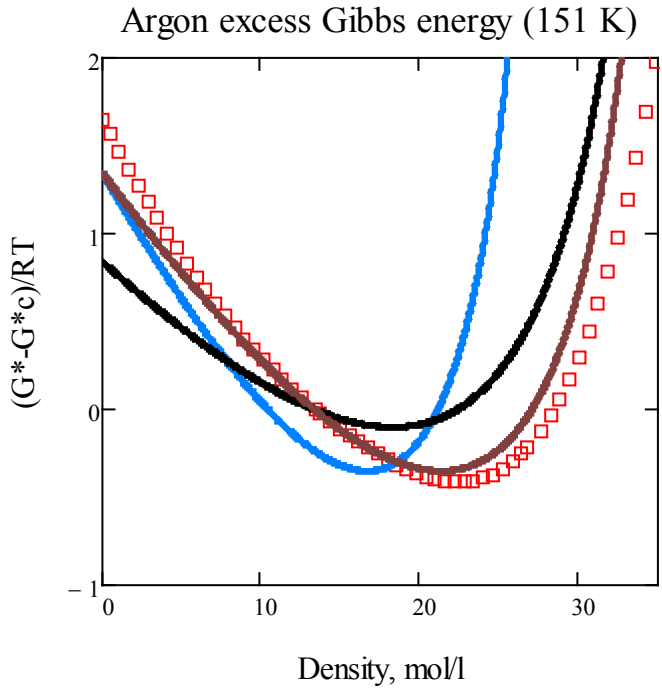

Fig. 2. Excess Gibbs energy $G^*$ of argon relative to $G_c^* = G^*|_{T_c,P_c}$ for the critical isotherm obtained from EOS [14] (red squares) as compared to the prediction of the van der Waals equation of state along critical isotherm: the solid blue line corresponds to $a = 27R^2T_c^2/64p_c = 1.337 \text{ atm} \cdot \text{l}^2 \cdot \text{mol}^{-2}$ and $b = RT_c/8p_c = 0.302 \text{ l} \cdot \text{mol}^{-1}$; the solid black line corresponds to $a = 3V_c^2 p_c = 2.43 \text{ atm} \cdot \text{l}^2 \cdot \text{mol}^{-2}$ and $b = V_c/3 = 0.025 \text{ l} \cdot \text{mol}^{-1}$; and the solid brown line corresponds to $a = 9RT_cV_c/8 = 1.045 \text{ atm} \cdot \text{l}^2 \cdot \text{mol}^{-2}$ and $b = V_c/3 = 0.025 \text{ l} \cdot \text{mol}^{-1}$.

23. As one can see from Eqs. 4-5 [1], Fig. 1 [1] and Fig. 3, the difference between the densities of liquid and gas coexisting in the phase equilibrium vanishes when $y = 0$. So, the statements in [2] that "The coexistence density difference function of $y$, $\Delta F(y) = b(\rho_{liq} - \rho_{gas})_{coex}$ must go to zero when $y = 1$. Plotting $F_{gas}(y)$ and $F_{liq}(y)$ against $y$, and finding that they have the singular value $\rho_c b = 1/3$ when $y = 1$ does not prove anything; there is no basis for assertion 2 above" are incorrect.

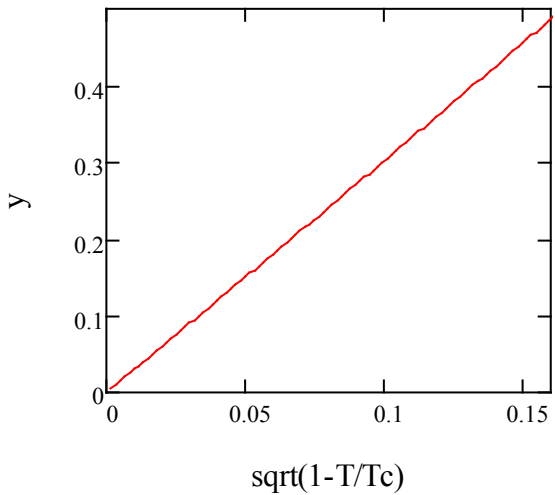

Fig. 3. The dependence of the coexistence density functional of the van der Waals fluid on $\sqrt{1 - T/T_c}$ obtained from Eq. 6 [1].

24. As one can see, $V \sim 0.05 \text{ l} \cdot \text{mol}^{-1}$ corresponds to $\rho \sim 20 \text{ mol} \cdot \text{l}^{-1}$. So, $V > b = 0.0320 \text{ l} \cdot \text{mol}^{-1}$. Therefore, the assertion in [2] that "The absurd minimum $G^*$ at

$\rho \sim 20 \, \text{mol} \cdot \text{l}^{-1}$ and subsequent increase for the hypothetical van der Waals liquid are consequences of $V < b$ in Eq. 1 at this density" is incorrect.

25. One can conclude using Eq. 6 [1] that the inequalities $0 \leq y \leq 0.5$ which are valid for Fig. 2 [1] correspond to the temperature interval $147 \, K \leq T \leq 151 \, K$ for $a = 1.337 \, \text{atm} \cdot \text{l}^2 \cdot \text{mol}^{-2}$ and $b = 0.0320 \, \text{l} \cdot \text{mol}^{-1}$ used in [2]. Therefore, the assertion in [2] that "the rigidity is related to the two reduced derivatives introduced in Ref. [4] and plotted against y for a very narrow ($<1 \, K$) near-critical range in figure 2 of [4]" is incorrect.

26. There exists the method for direct experimental measure of a critical density – the disappearance of the meniscus method which gives a high precision of the critical density determination (±0.02%) [6,20-30]. So, the statement in [2] that "No research in history has reported the direct experimental measurement of "a critical density" is incorrect.

27. The parameter $y$ was used earlier in [9-11] which were cited in [1]. As one can see from the definition of the parameter $y = 1/2 \cdot \ln[[1/bn_V(T) - 1]/[1/bn_L(T) - 1]]$ [1], it depends on the constant $b$ and saturation densities of the liquid ($n_L(T)$) and vapor ($n_V(T)$) of the van der Waals fluid. $n_L(T)$ and $n_V(T)$ are defined from the thermodynamic conditions of the phase equilibrium of the van der Waals fluid which are defined by the van der Walls equation of state (see Eqs. 1-3 [1]). Hence, $y$ depends on the constant $a$ of the van der Waals equation of state too. So, the parameter $y$ is not defined independently of the van der Waals equation functionally. Therefore, the statement in [2] that "Umirzakov [4] proposes a new coexistence state function $y(T)$, of subcritical gas and liquid densities ($\rho_{gas}$ and $\rho_{liq}$), respectively, which is defined independently of van der Waals equation functionally" is incorrect.

28. According to the parametric solution [1] of the equations corresponding to the liquid-vapor phase equilibrium of the van der Waals fluid the temperature dependence of the parameter, $y$ is defined from Eq. 6 [1], which is:

$$bkT/a = F(y(T)), \tag{4}$$

then the temperature dependencies of the densities of liquid ($n_L(T)$) and vapor ($n_V(T)$) of the van der Waals fluid are defined from Eqs. 4-5 [1] which are

$$bn_L(T) = F_L(y(T)), \tag{5}$$
$$bn_V(T) = F_V(y(T)). \tag{6}$$

A correct comparison of the phase equilibrium line of the van der Waals fluid with that of real fluid implies the definition of $n_L(T)$ and $n_V(T)$ from the above Eqs. 4-6.

As one can see, the temperature dependence of the parameter $y$ is defined by Woodcock [2] from

$$y_W(T) = 0.5 \cdot \ln\{[1/b\rho_{gas}(T) - 1]/[1/b\rho_{liq}(T) - 1]\}, \tag{7}$$

where $\rho_{gas}(T)$ and $\rho_{liq}(T)$ are the densities of the liquid and vapor of the real fluid (argon) coexisting in phase equilibrium; then he defines some functions $\rho_{L,W}(T)$ and $\rho_{G,W}(T)$ from

$$b\rho_{L,W}(T) = F_L(y_W(T)), \quad (8)$$
$$b\rho_{V,W}(T) = F_V(y_W(T)). \quad (9)$$

It is clear from Eqs. 7-9 that: $y_W(T)$ is not the parameter of the van der Waals fluid, so, $y_W(T) \ne y(T)$; $\rho_{L,W}(T)$ and $\rho_{G,W}(T)$ are not the densities of the liquid and vapor of the van der Waals fluid; and $\rho_{L,W}(T) \ne n_L(T)$ and $\rho_{V,W}(T) \ne n_V(T)$ because $y(T) \ne y_W(T)$. It is easy to see that: the values of $y_W(T)$ at critical temperature are presented in the last column of Table 1 [2]; the dependence $y_W(T)$ is presented in Fig. 3 [2]; the functions $F_L(y_W(T))$ and $F_V(y_W(T))$ are presented by the solid blue lines in fig. 4 [2]; the rigidity $\omega$ which is defined from Eq. 8 [2] using $y_W(T)$ is presented by the solid blue line in Fig. 5 [2]. So, the comparisons made by using the last column of Table 1 [2] and Figs. 3-5 [2] do not concern the van der Waals fluid. Therefore, the conclusions in [2] made by using $y_W(T)$ and based on the comparisons do not concern the <u>Assertion 2</u> and the conclusions in [1]. It is evident that the comparisons do not concern also the <u>Assertion 1</u>. It is also clear that the comparisons and all conclusions in [2] such as:
- "These results for the density difference coexistence state functional summarized in Table 1 show that the function $y(T)$ does not go to zero at $T_c$ as claimed "proven" in Ref. [4], but remains finite";
- " A plot of Eq. 7, i.e., $y(T)$, for argon as the experimental fluid, is seen to behave quadratically, with near-perfect regression ($R = 0.9999$), and interpolates to a constant nonzero value at $T_c$ as shown in Fig. 3. There is no evidence, experimental or otherwise, nor any good theoretical reason to believe any departure from this result within a tiny fraction of 1 degree K below $T_c$. The density difference $\Delta F(y) = F_{liq}(y) - F_{gas}(y)$ (see figure 1 of Ref. [4]) does not go to zero $T_c$ in the case of a real fluid";
- "the fact that "$\Delta F$", "$H$" and "$G$" go to zero at $y = 1$ for both coexisting gas and liquid in figures 1 and 2 of Ref. [4], respectively, does not prove anything about criticality of real fluids"; and
- "Figure 2 in Ref. [4], showing that these two derivatives go to zero when $y = 1$, does not prove anything because ω and its density derivatives all go to zero for real gas and liquid states at $T_c$, $p_c$. This is illustrated in Fig. 5 for the behavior of the rigidity of argon along the critical isotherm, compared to the prediction of van der Waals equation" based on $y_W(T)$ have no sense.

29. Fig. 4 demonstrates that the van der Waals equation of state (Eq. 1) quantitatively describes the experimental dependencies [31] of the saturated pressure of argon on temperature and reduced (to critical) vapor density near critical point, and it can describe qualitatively the reduced vapor and liquid densities of argon on temperature when the parameters $a$ and $b$ are defined from $(T_c, p_c)$.

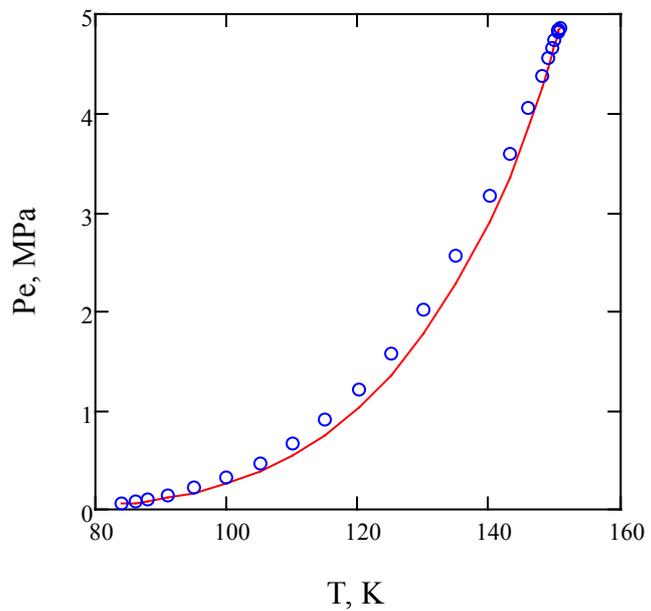

a) saturation pressure

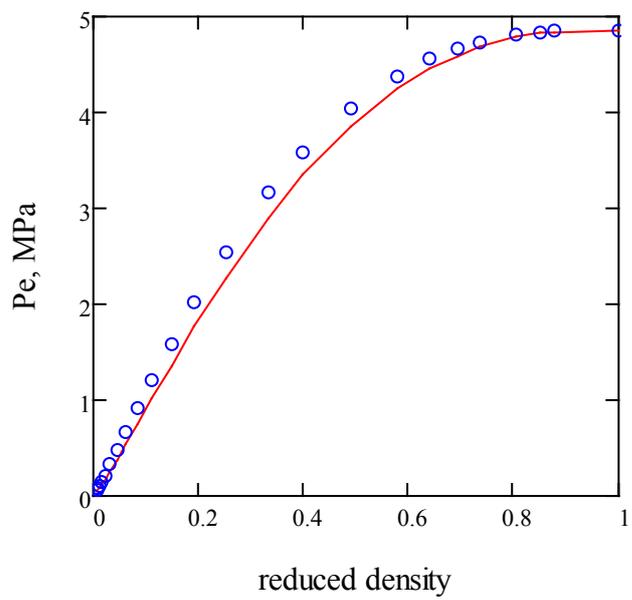

b) saturation pressure

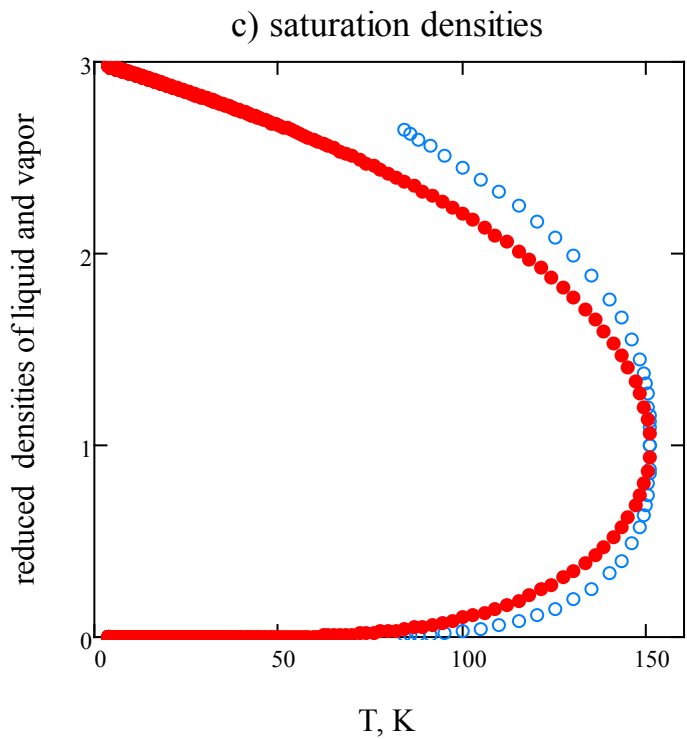

c) saturation densities

Fig. 4. The dependencies of the saturation pressures of argon [31] (blue circles) and VDW-fluid [1] (solid line) on temperature a) and reduced density b). c) The temperature dependencies of the reduced experimental coexistence densities of argon (blue circles) [31] and VDW-fluid [1] (red filled circles). $a = 27R^2T_c^2/64p_c = 1.337 \text{ atm} \cdot \text{l}^2 \cdot \text{mol}^{-2}$ and $b = RT_c/8p_c = 0.302 \text{ l} \cdot \text{mol}^{-1}$.

30. Fig. 5 shows that the van der Waals equation of state (Eq. 1) describes quantitatively the experimental dependencies [31] of the saturated pressure of argon on density and reduced (to critical) temperature near critical point, and it can describe qualitatively the vapor and liquid densities of argon on reduced temperature when the parameters $a$ and $b$ are defined from $(V_c, p_c)$.

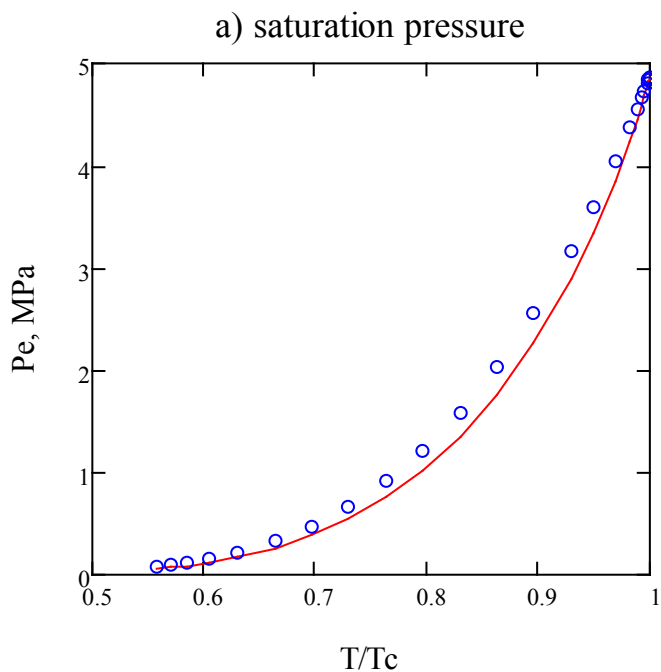

a) saturation pressure

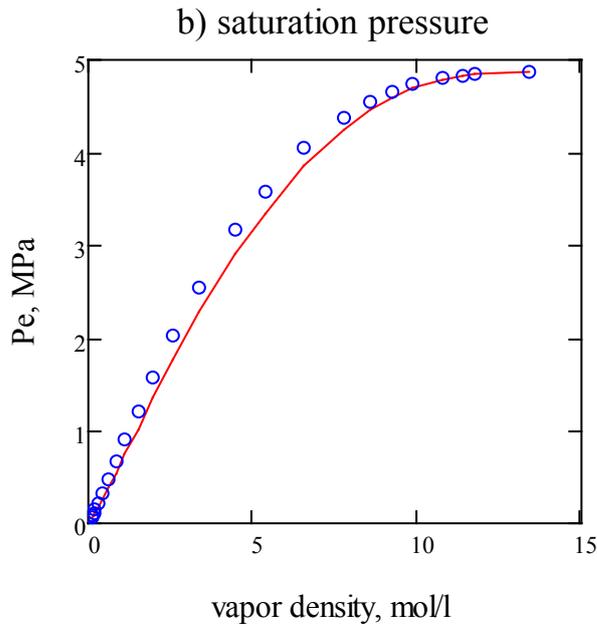

b) saturation pressure

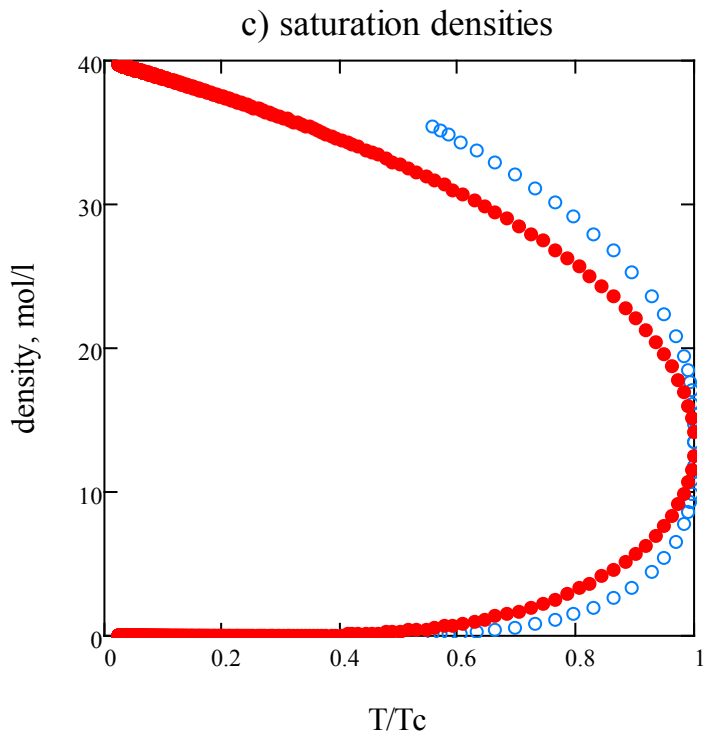

c) saturation densities

Fig. 5. The dependencies of the saturation pressures of argon [31] (blue circles) and VDW-fluid [1] (solid line) on reduced temperature a) and vapor density b). c) The dependencies of coexistence densities of argon (blue circles) [31] and VDW-fluid [1] (red filled circles) on the reduced temperature.
$a = 3V_c^2 p_c = 2.43 \, \text{atm} \cdot \text{l}^2 \cdot \text{mol}^{-2}$ and $b = V_c/3 = 0.025 \, \text{l} \cdot \text{mol}^{-1}$

31. Fig. 6 shows that the van der Waals equation of state (Eq. 1) can describe quantitatively the experimental dependencies [31] of the reduced (to critical) saturated pressure of argon on density and temperature near critical point, and it describes qualitatively the vapor and liquid densities of argon on temperature when the parameters $a$ and $b$ are defined from $(V_c, T_c)$. One can see from Eqs. 5-6 [1] and Fig. 1 [1] that the saturated vapor density of the van der Waals fluid is non-negative for any value of the temperature. Figs. 4 c), 5 c) and 6 c) show the same. So, in contrast to the Capture of Fig. 4 [2], the van der Waals equation of state is not absurd for temperatures below 110 $K$.

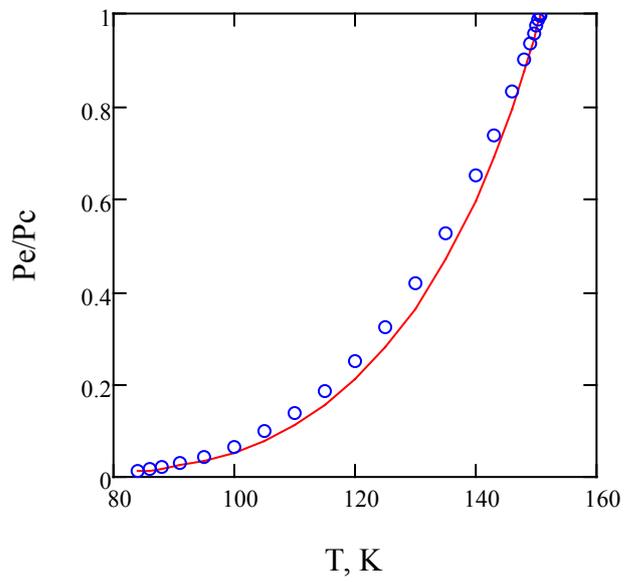

a) Reduced saturation pressure

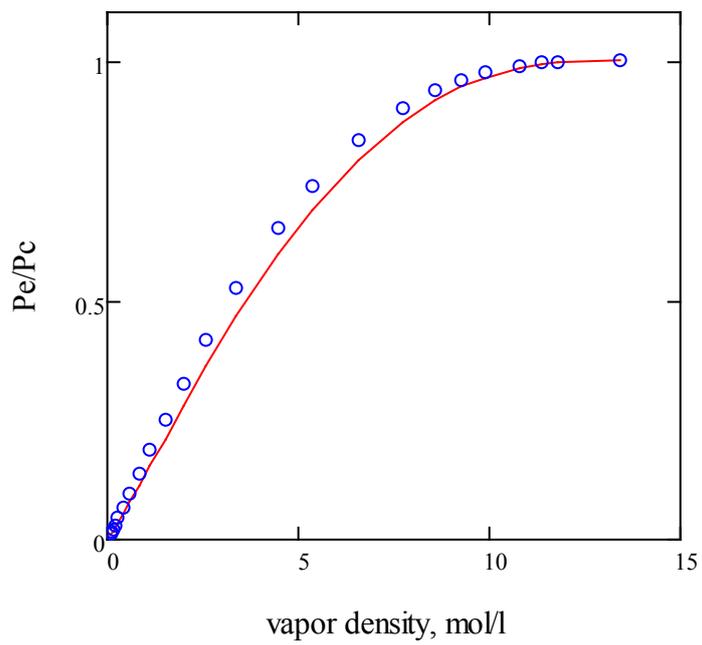

b) reduced saturation pressure

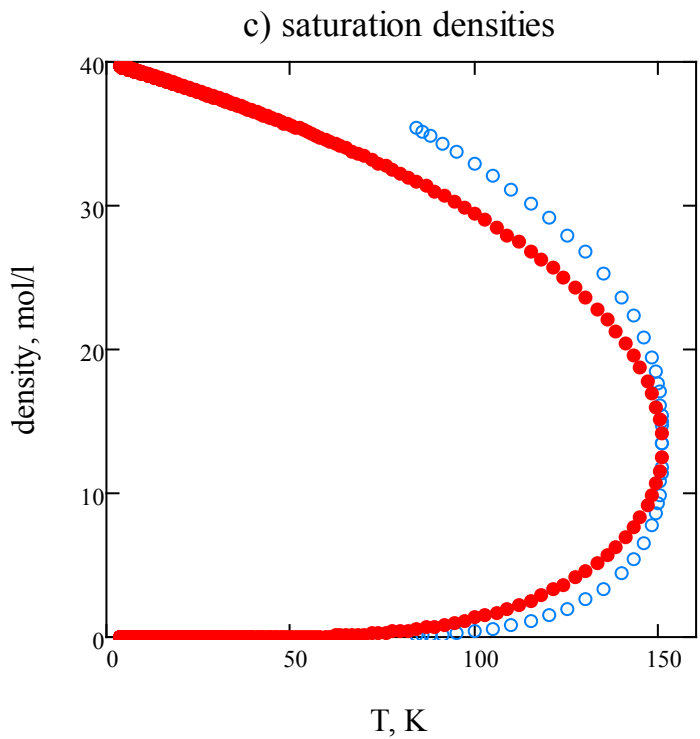

c) saturation densities

Fig. 6. The dependencies of the reduced saturation pressures of argon [31] (blue circles) and VDW-fluid [1] (solid line) on temperature a) and vapor density b). c) The dependencies of coexistence densities of argon (blue circles) [31] and VDW-fluid [1] (red filled circles) on temperature. $a = 9RT_cV_c/8 = 1.045 \; \text{atm} \cdot \text{l}^2 \cdot \text{mol}^{-2}$ and $b = V_c/3 = 0.025 \; \text{l} \cdot \text{mol}^{-1}$

32. One can see from Fig. 7 a) that the van der Waals equation of state can describe qualitatively the density dependence of rigidity of argon at critical temperature. Fig. 7 b) shows that the rigidity of argon is not equal to zero except the critical density.

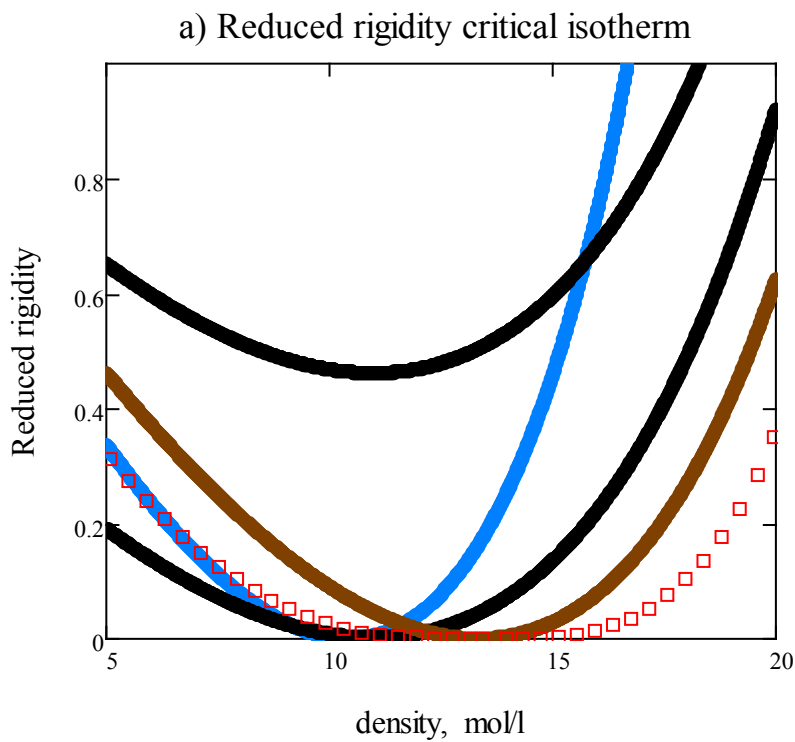

a) Reduced rigidity critical isotherm

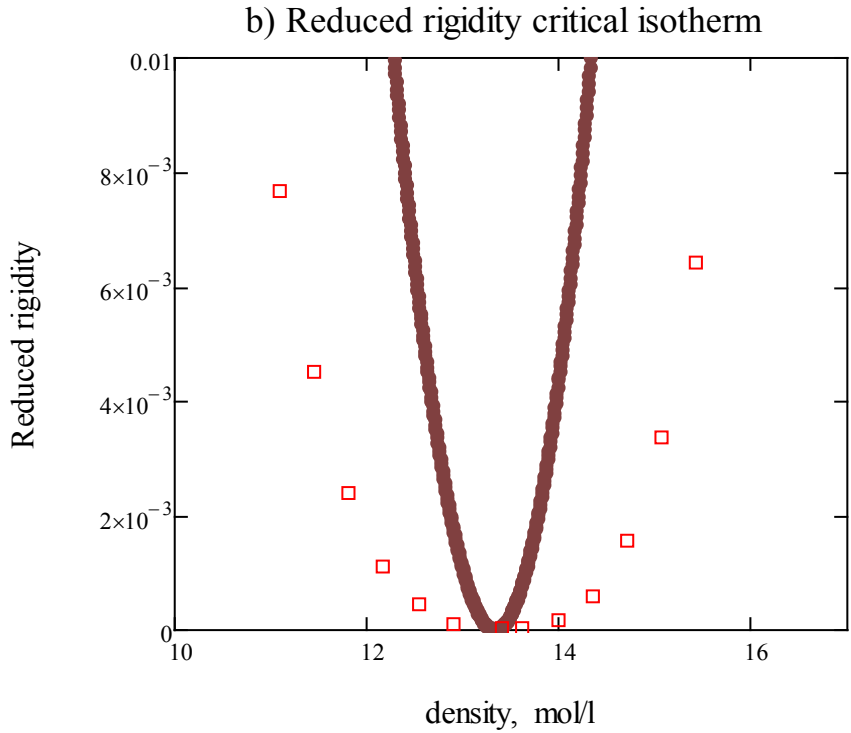

Fig. 7. Rigidity $\omega$ of argon defined from EOS [14] (red squares) compared with the prediction of the van der Waals equation of state along critical isotherm in the vicinity of the critical point: a) solid blue line corresponds to $a = 27R^2T_c^2/64p_c = 1.337 \text{ atm} \cdot l^2 \cdot \text{mol}^{-2}$ and $b = RT_c/8p_c = 0.302 \text{ l} \cdot \text{mol}^{-1}$; the upper solid black line corresponds to $a = 3V_c^2 p_c = 2.43 \text{ atm} \cdot l^2 \cdot \text{mol}^{-2}$ and $b = V_c/3 = 0.025 \text{ l} \cdot \text{mol}^{-1}$, the lower solid black line corresponds to the shift down of rigidity by $0.46$; solid brown line corresponds to $a = 9RT_cV_c/8 = 1.045 \text{ atm} \cdot l^2 \cdot \text{mol}^{-2}$ and $b = V_c/3 = 0.025 \text{ l} \cdot \text{mol}^{-1}$; and b) shows that the rigidity of argon is not equal to zero except the critical density.

33. According to the scaling theory which has a strong physical basis and quantitatively describes the thermodynamic properties of fluid near critical point [6,12], the density difference between gas and liquid vanishes at critical point and the temperature dependencies of saturation densities of the gas and liquid near critical point are determined by the equations

$$\rho_{liq}/\rho_c = 1 + c(T_c - T)^\beta, \quad \rho_{gas}/\rho_c = 1 - c(T_c - T)^\beta, \qquad (10)$$

where $c > 0$, $\beta > 0$ and $\beta \neq 1/2$. One can conclude from Eqs. 7 and 10 that the parameter $y$ vanishes at critical temperature. Therefore, the assertion in [2] that "A plot of … $y(T)$ for argon as the experimental fluid, is seen to behave quadratically, with near-perfect regression ($R = 0.9999$), and interpolates to a constant nonzero value at $T_c$ as shown in Fig.3. There is no evidence, experimental or otherwise, nor any good theoretical reason to believe any departure from this result within a tiny fraction of 1 degree $K$ below $T_c$" is incorrect.

34. One can see from comparison of contents of [1] and [2] that [2] does not include the proofs of the incorrectness of the assertions and conclusions made in [1]. One can also see the same from the comments presented above.

**Conclusions**

One can conclude from above considerations that there are a great number of incorrect equations and mathematical and logical errors in [2].

We have shown that: the dependencies for the isochoric heat capacity, excess Gibbs energy and coexisting difference functional of argon, and coexisting densities of liquid and vapor of the van der Waals fluid presented in all Figures in the paper [2] are incorrect; Table 1 [2] includes incorrect values of coexisting difference functional; [2] includes many incorrect equations, mathematical and logical errors, incorrect comparisons and incorrect assertions concerning the temperature dependences of the isochoric heat capacity and entropy of the real fluids; most of the conclusions in [2] are based on the above errors, incorrect data, incorrect comparisons and incorrect dependences. Therefore, the conclusions in [2] are not valid.

It is also shown that: the van der Waals equation of state quantitatively can describe the dependencies of saturation pressure on saturated vapor density and temperature near critical point; and the equation of state can describe qualitatively the excess Gibbs energy, rigidity and densities of coexisting liquid and vapor of argon (real fluid), including the region near critical point.